# A STUDY ON THE CHARACTERISTICS OF INTRODUCED PEARL MULLET POPULATION (Chalcalburnus tarichi, Pallas, 1811) IN LAKE ERÇEK

Sedat GÜNDOĞDU
Cukurova University, Faculty of Fisheries, Department of Fisheries 01330, Adana, TURKEY
e-mail: sgundogdu@cu.edu.tr

**Abstract**

This study examines features such as the population structure, growth and reproduction of the pearl mullet captured in the Lake Ercek. Between the dates January 2008 and January 2010 a total number of 527 individual were sampled through using trammel net and mesh size of 20,22 and 24 mm of fish net. The fork length, total and gonad weight, sex and fecundity of the pearl mullet which were captured has been identified; their age determination has been done through their operculum and otolith; their condition factor and gonadosomatic index value has been estimated.

**Key Words:** *Chalcalburnus tarichi*, fish biology, pearl mullet, population dynamic, Lake Erçek

**Introduction**

Pearl Mullet's which are a member of the Cyprinideae family is an endemic fish species which lives in the Lake Van Basin. When it is considered from the inland water fish production in Turkey, after the carp it ranks and being second by constituting $1/4^{th}$ of the inland water fish production. The Pearl Mullet's which is in the state of being the source of income for 14 thousand people around the Lake Van,, also contributes a great deal to the country's and region's economy. The studies done in the name of such an important species like the Pearl Mullet, generally takes place in the Lake Van. The first research for Pearl Mullet's were done by Pallas (1811). This taxonomic featured study later on was followed by Günther (1868), Deyrolle (1872), Ladiges (1960), Berg (1964), Kuru (1975), and Danulat and Selçuk (1992). The first researches' which examine the biological features was done by Akgül(1989). Later on this study was followed by studies done by Özdemir(1982), Özdemir *et. al.*(1985), Akyurt *et. al.*(1985), Küçüköner (1990), Danulat and Kempe (1992), Odabaşoğlu (1993), Şen (1993), Çetinkaya *et. al.*(1995), Arabacı (1995), Elp (1996), Sarı (1997), Arabacı *et al.* 2004), Arabacı and Sarı (2004) and Sarı (2009). The Pear Mullet which was trying to be made into a plus value for the economy, has been taken in time from it's natural environment that is the Lake Van and which has been introduced to other lakes in the basin, has also been introduced in the Lake Erçek in 1985. As for the research which is done on the population of Pearl Mullet in Lake Erçek is quite limited. Apart from the study done by Sarı and İpek(1998), no other research was encountered.

This study is the first one to be done on the population of the Peal Mullet in Lake Erçek. In the study some biological features of the Pearl Mullet in Lake Erçek and the population parameters were examined, the examined parameters has been compared with previously done studies on the population of Pearl Mullet in Lake Erçek.
.

**Material and Method**

Lake Erçek which is the study zone, is a lake that takes place 30 km south of Lake Van. It's waters are alkaline featured and it's pH value is in between 10.75 and 9.40. The annual temperature variation is in between 2-23°C (Yıldız, 1997). It's surface area is 114 km² and its surface altitude is 1810.52 m. Lake Erçek is the biggest lake after Lake Van in the basin. Its deepest point is 40 m while its average deep is 18.45 m (Sarı and İpek, 1998). The only one stream which inflow the lake is the Memedik Creek that falls from the south of the lake. Naturally, no naturaly fish species can live in Lake Erçek.

But in the year 1985, Pearl Mullet adults and fingerlings caught from Lake Van were introduced in the Lake by Provincial Directorate of Agriculture of Van.

The Pearl Mullet diadromous which is the study material is a fish specie. Its body shape is fusiform, its head lenght is approximately $1/6^{th}$ of the whole lenght. The mouth is horizontal and quite skewed and doesn't open much. As for the gill rakers which is generally 17-21 in number, it can also change between 14-27. Generally while the ruling colour is bright silvery, the back is dark greyish green or dark grey, the ventral area is a bright silvery colour (Geldiay and Balık, 1988). The Pearl Mullet immigrate altogether in the months of May-June to the fresh waters inorder to reproduce.

**Figure 1. The Location of Lake Erçek in Turkey and Lake Van Basin (Sarı and İpek, 1998)**

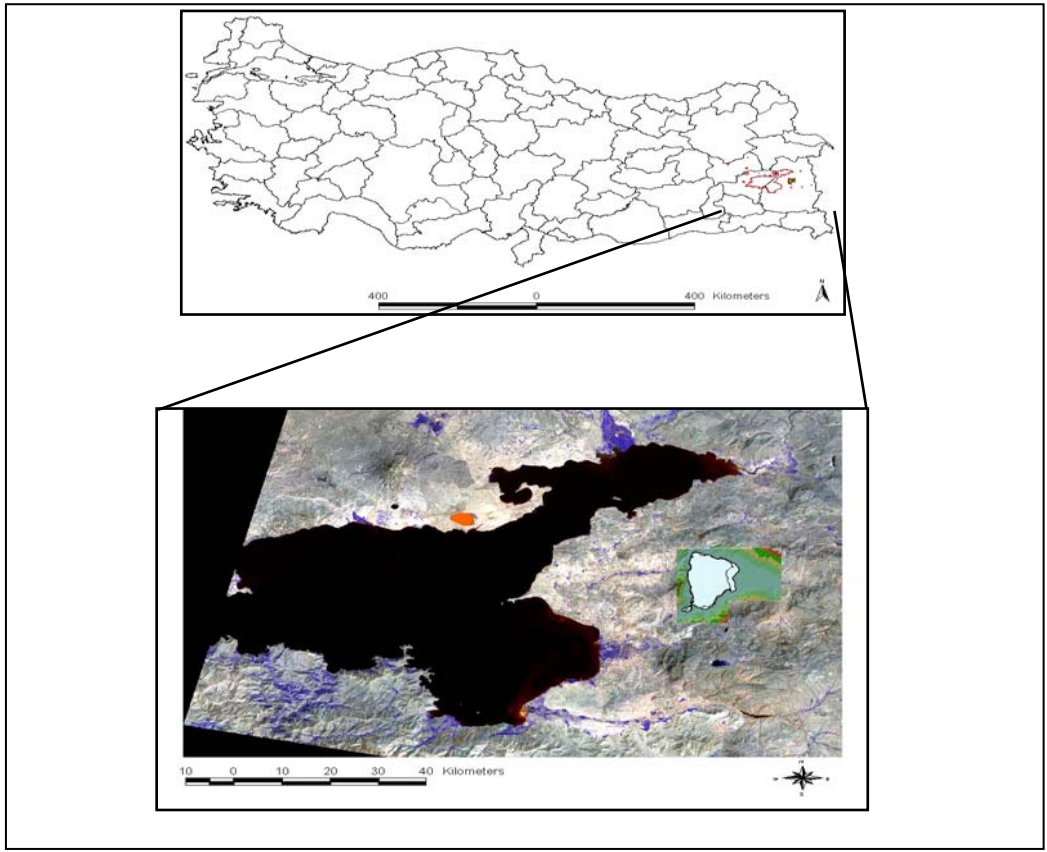

The study which has been carried out between the months of January 2008- January 2010, has sampled 527 Pearl Mullet 12 times by using 20 mm, 22 mm and 24 mm trammel nets. As for their reproduction period, it has been sampled with the help of casting nets in the river mouths and outfalls. Throughout the study period, the fork length and the weight of the sampled Pearl Mullet's, has been measured on a monthly basis, their sextined inheritance investment has been done and their gonad weight and their fecundities has been calculated. In the samples which were transfered to the laboratory, the fork length measurements, given by Lagler (1978), were both measured and evaluated with 0.5 cm sensitivity and their weight with 0.01 cm sensitivity. The age determination was done through operculum and otoliths, in accordance by Chilton and Beamish (1982). The sex determination was done given by Lagler (1978). The lenght-weight relationship was done through the given by Avşar (2005), as for the von Bertalanffy growth equations constants have been calculated in accordance by Sparre *et. al.* (1989). The condition factor has been calculated with the help of equality which is also known as the Fulton's condition factor; fecundity has been calculated gravimetrically, through getting subsamples from samples as it was informed by Avşar (2005) and the fecundity-lenght relationship was calculated in accordance to Avşar (2005). It has been utilised from Düzgüneş (1983) in evaluating the study results statistically. As for doing the statistical operations, SPSS v16 program were used.

**Results**

In the research, 527 Pearl Mullet which is distributed in the ages between II-IIV has been seen and the samples distributional percentage according to ages are respectively calculated as %4.3, %29.1, %22.7, %21.7, %15.3 and %6.9. Among the samples, while the 3 year old age group has been found in the highest percentage with %29.1, the 2 year old age group has been found in the lowest percentage with %4.3.

It has been confirmed that 250 of the total number of 527 Pearl Mullets which have been examined and which their sexual investments were completed, were male and 277 of it were female. The male:female percentage of the samples have been approximately calculated as 1:1. Through deriving benefit from the measured individual lenght and weight data the constant lenght-weight relationship acquired for all individuals are found as; 0.02(a), 2.83(b), 0.93(r), for males are; 0.014(a), 2.94(b), 0.91(r), for females are; 0.20(a), 2.83(b), 0.93(r). As it is seen, while it is close to 3, the "b" constant of the lenght-weight relationship, even if just a drop, is found to be smaller than 3. In this case, it can be said that Pearl Mullet's of Lake Erçek show a negative allometric growth. The calculated average condition factors are respectively calculated as; for all individuals; $1.255\pm0.005$, for males; $1.269\pm0.007$ and as for females; $1.240\pm0.008$. When the condition factor value is taken in to consideration, it can be said that in total, males posses a higher condition than females. The von Bertalanoffy growth constants were estimated as Lson…, tsıfır…., K…. and Wson…. The von Bertalanffy growth equation for all individuals, males and females have been found as; $L_t=39.5229*(1-e^{-0.089(t-(-5.096))})$ and $W_t=699.2525*(1-e^{-0.089(t+5.096)})^{2.844}$. the lenght values which have been measured and calculated for all males, females and all individuals are given in Chart 1 and figure 2.

**Chart 1. The Measured and Calculated Lenghts for Male, Female and Their Pooled Data Age Groups**

|  | Ages | | | | | | |
|---|---|---|---|---|---|---|---|
|  | I | II | III | IV | V | VI | VII |
| **Male** | | | | | | | |
| Observed | * | 17.48±0.2 | 20.33±0.06 | 22.26±0.06 | 23.40±0.06 | 24.15±0.1 | 26.68±0.3 |
| Calculated | 18.38 | 20.76 | 22.73 | 24.36 | 25.71 | 26.83 | 27.76 |
| **Female** | | | | | | | |
| Observed | * | 18.00±0.7 | 20.60±0.07 | 22.33±0.07 | 23.78±0.07 | 24.70±0.07 | 26.73±0.1 |
| Calculated | 18.72 | 20.93 | 22.82 | 24.43 | 25.80 | 26.96 | 27.96 |
| **Pooled Data** | | | | | | | |
| Observed | * | 18.01±0.8 | 20.08±0.05 | 21.60±0.04 | 22.86±0.05 | 24.18±0.06 | 25.78±0.1 |
| Calculated | 16.49 | 18.45 | 20.25 | 21.88 | 23.38 | 24.75 | 26.00 |

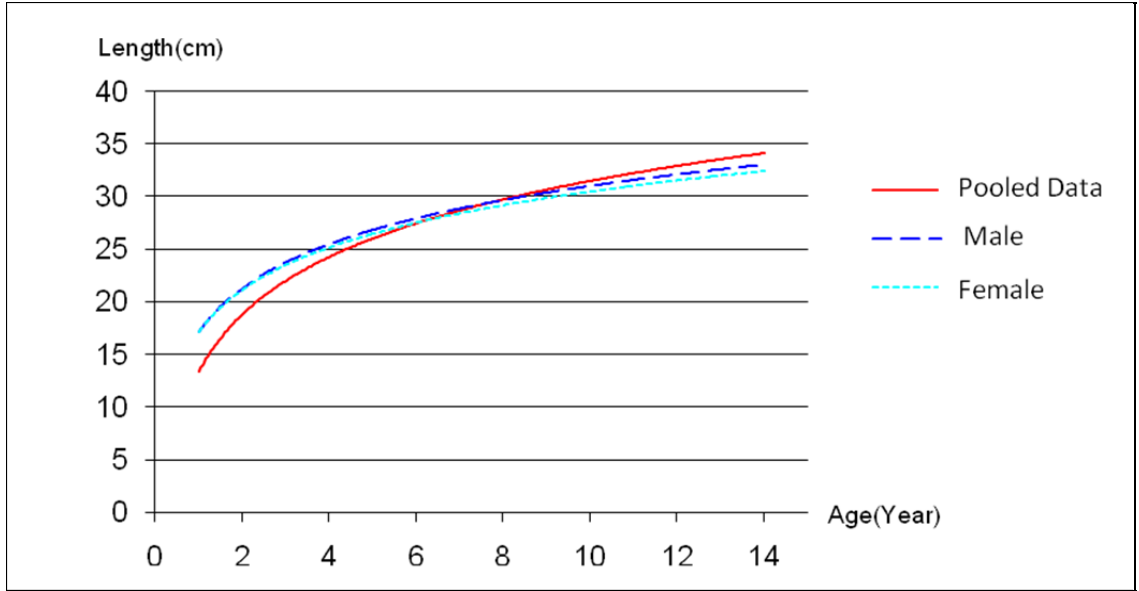

**Figure 2. The Calculated Lenght for Males, Females and Their Pooled Data.**

Sedat, bu grafikler üzerinde gerçek datalara ait dağılım noktaları gözüksün. Grafik üzerinde mutlaka tsifir değeri de gözüksün.

In the research, it has been calculated as $W_\infty$=699.25g. the measured and calculated weights are given in Chart 2. and Figure 3.

**Chart 2. The Measured and Calculated Weights for Male, Female and Their Pooled Data**

| Age Groups | Ages | | | | | | |
|---|---|---|---|---|---|---|---|
| | I | II | III | IV | V | VI | VII |
| **Male** | | | | | | | |
| Observed | * | 76.30±2.8 | 99.93±1.5 | 124.88±1.7 | 148.66±2.0 | 169.41±3.4 | 209.32±10.0 |
| Calculated | 77.07 | 110.27 | 143.98 | 176.55 | 206.94 | 234.57 | 259.22 |
| **Female** | | | | | | | |
| Observed | * | 67.26±4.1 | 102.79±1.2 | 129.29±1.6 | 154.77±1.7 | 178.21±2.2 | 211.17±3.9 |
| Calculated | 85.41 | 117.25 | 149.76 | 181.67 | 212.11 | 240.53 | 266.61 |
| **Pooled Data** | | | | | | | |
| Observed | * | 73.67±1.7 | 101.06±1.1 | 125.19±1.2 | 152.17±1.3 | 175.60±1.9 | 217.82±3.7 |
| Calculated | 59.67 | 82.08 | 106.79 | 133.18 | 160.71 | 188.87 | 217.25 |

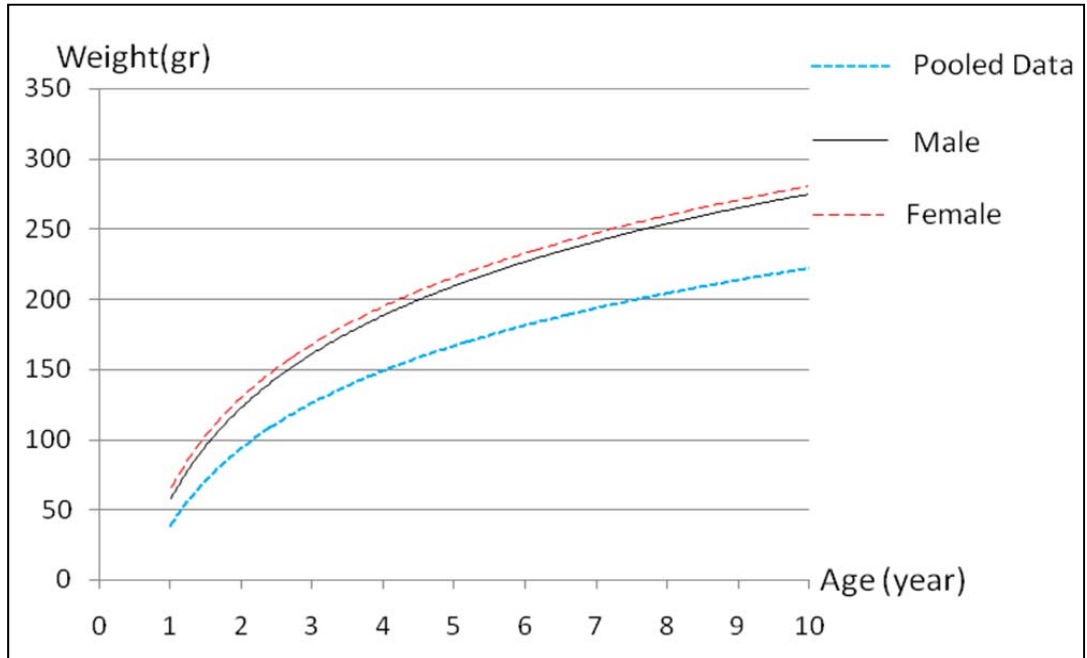
**Figure 3. The Calculated Weight for Males, Females and Their Pooled Data.**

The gonado-somatic index (GSI) values which were used to identify the reproduction time, were calculated separately in reference to months for males, females and all individuals. According to this the minimum values were confirmed in July while the maximum values were confirmed in April and when taking these data's into consideration, it can be understood that reproduction of the Pearl Mullet's of Lake Erçek starts in April and continues until July.

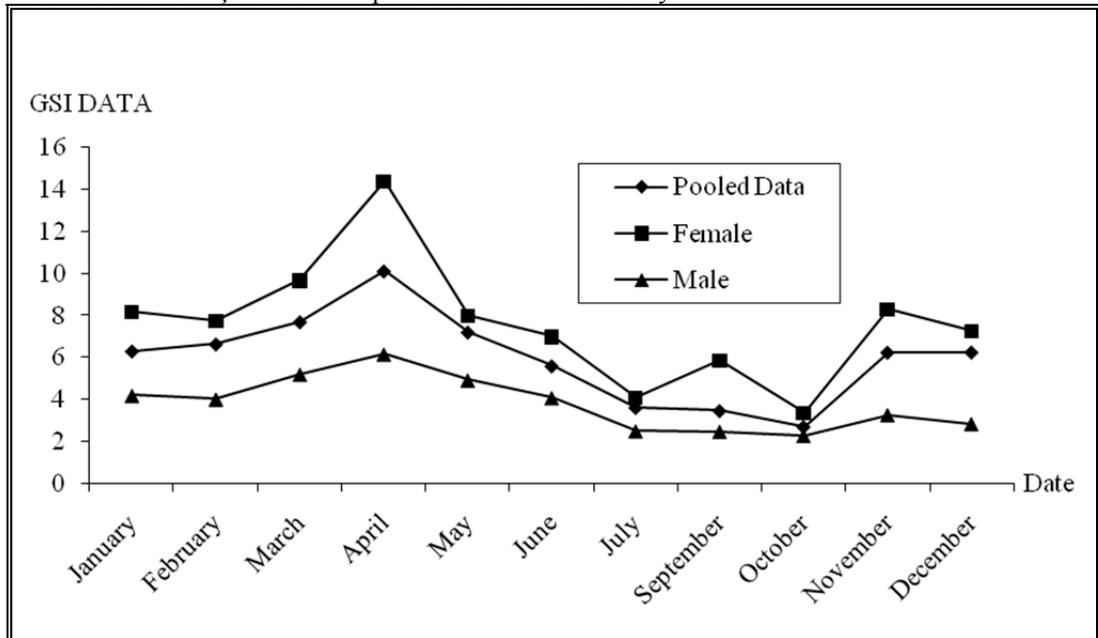
**Figure 4. The Whole Samples of the Sampled Pearl Mullet's, the Monthly Change of Male, Female and Their Pooled Data GSI Values.**

The acquired fecundity values of 147 fish out of 527 caught ones, were calculated separately with respect to age. According to this, respectively, the average fecundity rates which were calculated

for each age group in between III-VII are 15798, 17924, 21434, 24206, 24053, 20109. It can be said that the fecundity begins to fall in VII oldest fish which are the biggest among the age groups.

**Discussion**

When the fish which were sampled with the research catching from Lake Erçek were examined, it was determined that the smallest age group was age II, as for the biggest age group, it was age VII. Among the sampled fish, the age I grouped fish wasn't into. It has been thought that the reason for not catching the age I group fish during the sampling, was because the mesh size of the used nets were too big for this age group. The results of the research resemble other studies which were done before. Sarı (1997), has calculated the Pearl Mullet's age which he sampled in Lake Van, for two different seasons and at each season he calculated it separately as winter period and reproduction period. In his research, Sarı(1997) has determined that the age range in the 1994-1995 catching season and in the 1995-1996 seasons, were in between ages II-VIII.

According to the obtained lenght values of the fish which caught from Lake Erçek, it has been estimated that the smallest lenght is 16 cm and the biggest lenght is 27 cm. Also all individual's average lenght values has been calculated as 21.985cm±0.0877. A great difference is observed between the identified minimum, maximum and average lenght values in this study and the ones which were identified in previous studies. In the research which he done ,Sarı(1997) has measured the lenght range of the 1994-95 fishing seasons and again at the same season's winter fishing while finding the average lenght as 18.05 cm, he has confirmed it as 16.83 cm for the reproduction period. He has confirmed that the lenght range in 1995-96 fishing season is in between 15-18 cm. While finding the average lenght as 17.15 cm in the same season's winter period; he has confirmed it as 15.44 cm in the reproduction period. According to Nikolskii (1980), the observable differences between the same species populations in different habitats can basically change according to the amount of the food sources, the biotic and abiotic features of the habitat.

According to the weight results of the sampled fish, it has been determined that the smallest weight is 44.9 g and the biggest weight is 265.2 g when all the meanweight results are taken into consideration, it has been calculated that the obtained average weight is 136.65±1.7. As it was in the case of the lenght values, the weight values were rather different from the previous studies as well. For example, Sarı (1997) has calculated the value of minimum and maximum weights for 1994-95 seasons as 50 -65 g, as for the average weight for the winter period it is 66.29 g and for the reproduction period he has calculated it as 53.06 g. It is thought that the observable difference between the previous studies and this study, like it was mentioned before, results from habitat differences.

It has been estimated that among the whole individual populations of the Pearl Mullet's of Lake Erçek, %47.5 of it are male and %52.2 of it are female. Like the case of the difference in lenght and weight values, there are also differences between the observed proportion of sexual inheritance in this study and the proportions of previous studies. While Sarı (1997) estimated the male female ratio as 1:1.65 for the 1994-95 season's winter period; he estimated this ratio as 1:0.57 for the reproduction period. He estimated the 1995-96 seasons winter period as 1:2.5, and for the reproduction period as 1:0.39. It is thought that the difference between this study and Sarı's (1997) study is due to habitat differences (Nikolskii, 1980).

The lenght-weight relationship is calculated for all individuals and it has been seen that the estimated (b) value is smaller than 3 (2.8447). Avşar (2005) has stated that the (b) value is an indication of the body shape of the fish. And this has been interpreted that the Pearl Mullet's of Lake Erçek show a negative allometric growth. When the estimated regretion constraints are compared to Sarı's (1997) findings, it has been calculated that the (a) value in this study is 0.0201, in Sarı's(1997) study it is 0.0654,the (b) value in this study is 2.844, in Sarı's(1997) study it is 2.276. this difference between (a) and (b) values has been interpreted that it arises from the access to the food source in Lake Erçek and Van, ovulation condition, heat and different ecological circumstances such as the features of a biotope (Ricker, 1975). It has been thought that the obtained differences between the functional regression constants can also arise from lenght measurement precision and the difference in the sampling periods (Sarı, 1997). It has been seen that the changes of the (b) value which is the functional regression constant, throughout a year is also relevant with such features of the Pearl Mullet's as feeding and reproduction. While this value has shown a decrease in the reproduction period, it has reached the highest level in the nutrition period.

The growth equation parameters of Pearl Mullet's Lake Erçek have been divided according to age groups and the values have been calculated from all these age groups without seasonal exceptions. In this study, the $L_\infty$ values has been calculated as 39.52 cm/yr. Sarı (1997), has calculated this value as 22.17 cm/yr in his study which he did. While the K value was calculated as 0.0894 in this study, Sarı (1997) has calculated this value as 0.3010, while the $t_0$ value was calculated as -5.09 in this study, Sarı (1997) has calculated it as -1.15. $t_0$ It has been observed that the obtained values in this study and the values which were calculated by Sarı (1997) show a difference and like it was mentioned before, this difference is attributed to the differences between habitats. Also it has been thought that the reason why $t_0$ value is high in this study is because age I group fishes could not sampled.

The ($W_\infty$) value has been calculated as 699.25g. In his study, Sarı(1997) has calculated the ($W_\infty$) value as 111 g. It is thought that, the weight value distinctions of the same species among different populations, arises from the different growth speed which they have in different habitats where populations live (Ricker, 1975). According to the results of the calculations it has been observed that the GSI value has reached it's highest peak in April and falls into decline until July. And this shows us that the reproduction period of Pearl Mullet's is in between the months of April-July (Avşar, 2005). Throughout the study period, the average condition value of the sampled Pearl Mullet's has been calculated in all individuals as 1.255±0.005, in males as 1.269±0.007 and for the females as 1.240±0.008 It has been thought that the condition value differences which is seen depending on sexual inheritance, is relevant with the fact that during the reproduction period the conditions of the females are low. Again, is has been explicated that the calculated condition value of Pearl Mullet's shows that they are nourished well. The average fecundities of the 147 fish whose fecundities were calculated, were found as 20109±606. While the calculated fecundity-lenght relationship was found exponential, the (b) value in this relationship was calculated as 2.4572. Avşar (2005), has stated that the (b) value in fecundity-lenght relationship of teleosteis is near 3.

**Acknowledgements**